\documentclass[useAMS,usenatbib]{mn2e}
\usepackage{times}
\usepackage{epsfig}
\usepackage{subfig}
\usepackage{natbib}
\usepackage{balance}
\usepackage{url} 
\usepackage{soul}
\usepackage{color}

\usepackage{amsmath, amssymb, graphicx, multirow, rotating, lscape}

\begin{document}

\title{ Multi-band variability in the blazar 3C 273 with XMM-Newton}

\author[Kalita et al.]{Nibedita\ Kalita$^{1,2}$, 
Alok C.\ Gupta$^{1}$\thanks{Email: acgupta30@gmail.com}, 
Paul J.\ Wiita$^{3}$, Jai\ Bhagwan$^{1,4}$, Kalpana\ Duorah$^{2}$ \\
\\      
\\
$^1$Aryabhatta Research Institute of Observational Sciences (ARIES), Manora Peak, Nainital, 263 002, India \\
$^2$Department of Physics, Gauhati University , Guwahati,781 014,India  \\      
$^3$Department of Physics, The College of New Jersey, P.O.\ Box 7718, Ewing, NJ 08628-0718, USA \\              
$^4$School of Studies in Physics \& Astrophysics, Pt Ravishankar Shukla University, Amanaka G.E.\ Road, Raipur 492010, India}

\date{}

\pagerange{\pageref{firstpage}--\pageref{lastpage}} \pubyear{2015}

\maketitle

\label{firstpage}

\begin{abstract}

We have undertaken a nearly simultaneous optical/UV and X-ray variability study of the flat spectrum radio 
quasar, 3C 273 using data available from the XMM$-$Newton satellite
mission from June 2000 to July 2012. Here we focus on the multi-wavelength flux variability on both intra-day 
and long time scales of this very well known radio-loud source. We found  high flux variability over long time 
scales in all bands for which observations were made. The optical/UV variability amplitude was more than twice 
than that in the X-ray bands. There is some frequency dependence of the variability in optical/UV bands in the 
sense that the variability amplitude increases with increasing frequency; however, the  X-ray emissions 
disagree with this trend as the variability amplitude decreases from soft to hard X-ray bands. On intra-day
time scales 3C 273 showed small amplitude variability in X-ray bands. A hardness ratio analysis in the 
X-ray regime indicates that the particle acceleration mechanism dominates the cooling mechanism during
most of the $\sim$12 year span of these observations.  

\end{abstract}


\section{Introduction}

The subclasses of radio-loud active galactic nuclei (AGN) that include BL Lac objects (BLLs) and flat spectrum
radio quasars (FSRQs) are collectively known as blazars. The optical spectra of BLLs show featureless continua 
whereas spectra of FSRQs have prominent emission lines.  Blazars show rapid flux variability across the complete 
electromagnetic (EM) spectrum, have high and variable polarization and  their  emission is predominantly non-thermal. 
Their radio structures are dominated by compact cores. Blazars also show superluminal motion due to Doppler boosting
and thus are understood to possess relativistic jets of highly collimated radiation and plasma emanating from the 
central region viewed at a small  ($\lesssim$ 10$^{\circ}$) angle to the line of sight of the observer (Urry \& 
Padovani 1995).  In the case of FSRQs  the accretion disc  component (big blue bump) can be seen in their multi-
wavelength spectral energy distributions (SEDs) on  occasions when the FSRQs are observed in their low flux state, 
whereas BLLs show weak or absent emission from the disc.

Blazars show flux variations in all bands of the EM spectrum on all possible timescales. Those fluctuations ranging 
from a few tens of minutes (and sometimes even more rapid) to less than a day in our observer's frame are often known 
as intra-day variability (IDV) (Wagner \& Witzel 1995). Variability with time scales of few days to weeks is commonly 
known as short term variability (STV), while variability on time scales ranging from months to the longest we have 
monitored (typically several years) is called long term variability (LTV) (e.g., Gupta et al.\ 2004).  
The  IDV in blazars  is the least well understood type of variations but it
can provide an important tool for learning about structures on small spatial scales and it also 
provides us with better understanding of the different radiation mechanisms that are important  in the emitting 
regions (e.g., Wagner \& Witzel 1995). Any characteristic time scale of variability can effectively provide 
us with information about the physical structure of the central region and  complex phenomena such as hot spots on 
accretion discs (e.g., Mangalam \& Wiita 1993). In blazars, it is widely believed that most variability originates 
mainly from  enhanced emission in a small region within the relativistic jets (e.g., Begelman et al.\ 2008). For 
example, excess emission could be triggered by dissipation inside a  blob of material emerging from at the jet base,
from shocks propagating along the jet, or  from instabilities within the jet itself. 

3C 273 ($\alpha_{2000.0}=12{\rm h} 29{\rm m} 06.7{\rm s}$, $\delta_{2000.0}=+02^{\circ} 03^{'} 09^{"}$, 
$z=$0.158339) was the first quasar 
discovered in the radio sky survey at 158 MHz that was included in the third Cambridge (3C) catalogue 
(Edge et al.\ 1959). In  early 1963, it was classified as a FSRQ (Schmidt 1963). Unsurprisingly, this relatively 
nearby blazar has been extensively studied on diverse time scales across the complete EM spectrum, with measurements
made in single bands and often as well as in multi-bands 
(Mantovani et al.\ 2000; Collmar et al.\ 2000; Sambruna et al.\ 2001; Greve et al.\ 2002; Kataoka et al.\ 
2002; Courvoisier et al.\ 2003; Jester et al.\ 2005; Attridge et al.\ 2005; Savolainen et al.\ 2006; McHardy 
et al.\ 2007; Soldi et al.\ 2008; Pacciani et al.\ 2009; Dai et al.\ 2009; Fan et al.\ 2009, 2014; Abdo et al.\
2010; and references therein). 

This source has been studied in optical bands since its discovery. Optical (V, R and I band) studies show 
the variable flux in the source changes with different amplitudes on different timescales revealing the nonlinear 
variability characteristic of blazars. It was also noticed that, at higher frequencies, the strength of the flux 
variations in 3C 273 are higher than in the lower frequencies (e.g., Fan et al.\ 2009). This type of frequency 
dependent variability is also valid for near-infrared bands (Wen et al.\ 2002).  While monitoring 
the source in optical bands from 1998 to 2008, Fan et al.\ (2014)  found that during this period the 
source did not show very significant IDV but it showed strong LTV.  Imaging data from {\it Hubble
Space Telescope} in the UV  show compatibility of far-UV with extrapolated 0.5-8 keV X-ray flux, thereby 
supporting a common origin of UV and X-ray emissions in the relativistic jet (Jester et al.\ 2007). 
Long-term optical observations in BVRI bands from 2003 to 2005 with the Yunnan National Astronomical 
Observatory (YNAO) 103 cm telescope and Shanghai Astronomical Observatory (SHAO) 156 cm telescope show 
that the source was in a rather steady flux state during this campaign (Dai et al.\ 2009).  The same study shows strong 
inter-band correlations of color index and magnitude in the optical regime, where the spectrum becomes flatter 
when the source brightens and softer when the source brightness decreases for both IDV and LTV, which is 
similar to spectral evolution of most BL Lacs. 

Soldi et al.\ (2008) used various ground and space based telescope data to study the variability in 
the source over almost the entire EM spectrum (i.e., radio to $\gamma-$rays) for about a decade. Their study 
revealed energy dependent variability time scales and variability amplitudes of 3C 273. The variability 
in X-ray bands indicate the presence of two different components, one of them which can be considered Seyfert-like 
while the other is blazar-like.  A contemporary study of the source with BeppoSAX and XMM-Newton interprets  the 
Seyfert-like component and soft excess as the result of a non-thermal flare-like coronal model (Pietrini et al.\ 2008). 
The hard X-ray variability correlated with  long term optical variability but not  to 
radio variability, which indicates that the emission mechanism for hard X-rays ($\geq$ 20 keV ) is not due 
to  electrons accelerated by the shock waves in the jet but probably arises from inverse Compton scattering of 
synchrotron optical photons emitted from a region near the base of the jet by the same electron 
population (Soldi et al.\ 2008). 

A soft X-ray study of 3C 273 by ROSAT showed short term variability of about 20 per cent over
$\sim 2$ days (Leach et al.\ 1995). RXTE observations of the source at hard X-ray bands found 
similarly modest variations over several days but variability up to a factor of four over four 
years   (Kataoka et al.\ 2002).  The high correlation  between X-ray (3$-$20 keV) and IR fluxes 
provides evidence for the origin of X-ray emission in 3C 273 being a result of the Compton scattering 
of low-energy seed photons (McHardy et al.\ 1999). However, the X-ray emission from 3C 273 shows 
different timing properties for variability below and above $\sim 20$ keV and this produces uncertainty 
about the origin of the hard X-ray emission, which was previously considered to be part of the jet emission 
(Soldi et al.\ 2008).They have studied fractional variability amplitude and characteristic time scale 
with structure function (SF) analysis using XMM-Newton data from 2000--2005. This analysis technique to study 
variability has been used frequently by the community. Turriziani (2011) \& Vagnetti et al.\ (2011) seperately 
performed SF to study the X-ray variability for this source with XMM-Newton data. The later work (Vagnetti et al.\ (2011))
found that the variability is anti-correlated with X-ray luminosity. During the period 2000--2009, Liu \& Zhang 
(2011) studied STV in soft and hard X-ray bands with the help of XMM-Newton satellite, and they reported that 
soft and hard X-rays come from different origins. Brightman \& Nandra (2012) used classical hardness ratio to 
find an effective colour--colour selection for obscured AGN from X-ray data using a sample of heavily absorbed 
active galactic nuclei (including 3C 273).

A recent study has indicated that the physical process initiating variability in the X-ray 
domain of 3C 273 is similar to that in Seyfert galaxies (McHardy 2013). 3C 273 has long been known to have 
a soft excess in the X-ray band at $\leq1$ keV (Turner et al.\ 1985).  Many studies have shown a soft 
excess at energies ranging from 0.1 keV $-$ 2 keV (Courvoisier et al.\ 1987; Page et al.\ 2004; T{\"u}rler et 
al.\ 2006; Chernyakova et al.\ 2007; Pietrini et al.\ 2008;  and references therein). It was also detected 
as one of the first extragalactic $\gamma-$ray sources by the COS-B satellite (Swanenburg et al.\ 1978).

A wavelet analysis of X-ray time series data of 3C 273 taken by XMM--Newton indicated a candidate 3.3 ks  quasi-periodic 
oscillation (QPO; Espaillat et al.\ 2008).  With the assumption that the QPO is generated near the last stable orbit in 
the accretion disk the central black hole mass is constrained to lie between $7.3\times10^6$ $\it M_{\odot}$ if it is  a
Schwarzschild BH and  $8.1\times 10^7$ $\it M_{\odot}$ for a maximally rotating Kerr black hole.  However, previous 
reverberation-mapping estimates gave much higher masses of $2.35\times10^8$$\it M_{\odot}$ (Kaspi et al.\ 2000) and 
$6.59\times10^9$$\it M_{\odot}$ (Paltani \& T\"{u}rler 2005) which would imply that this QPO is not due to the dynamical 
motion in the inner accretion disk, but perhaps due to oscillation modes in the accretion disc (Espaillat et al.\ 2008). 
Subsequent studies have, however, called into question the presence of a QPO in 3C 273 (Mohan et al.\ 2011; 
Gonz{\'a}lez-Mart{\'{\i}}n \& Vaughan 2012). 

Although the emission mechanism for blazars is not yet completely understood, it is clear that the lower energy photons 
arise from synchrotron emission and in many cases this emission extends into the soft part of the X-ray band.
Contributions from synchrotron self-Compton (SSC) and external Compton (EC) are the two main candidate mechanisms 
to dominate the harder part of X-ray emission and produce the $\gamma$-ray emission.  The harder X-ray emission also 
can have a contribution from the Compton scattering of disc photons by the hot corona sandwiching the disc (Sikora 1994, 
Ghisellini \& Tavecchio 2008). Studies have also shown strong correlation between IR and X-ray emission from time to time, 
in the sense that the IR leads the X-ray emission, supporting the SSC model (McHardy et al.\ 2007).

Multi-wavelength (optical, UV and X-ray) LTV and interband correlation of this source 
with XMM-Newton data available till date (2012) is done for the first time by us in the present work. LTV of 3C 273 in 
X-ray bands was studied previously in different time period by differet groups, but here we study LTV of this source 
for the longest period ever done in X-ray bands, both soft and hard bands using XMM-Newton data. We also study the 
LTV in optical and UV bands in the same period with Optical Monitor (OM) data, which has not done before. IDV studies 
on a few occasions are done in past but we have studied here IDV in X-rays with all the time series data available for 
the source till 2012.

In this paper, we study IDV and LTV of the FSRQ 3C 273. In the next section, we discuss XMM$-$Newton archival 
data. In section 3, the techniques we used for data reduction and light curve generation are given. In section 4, 
we report our results and in section 5, we present a discussion and the conclusions of the work.

\section{XMM$-$Newton Archival Data}

\subsection{X-ray Data}

3C 273 has been frequently observed with different telescopes over different frequency ranges. The European X-ray satellite 
mission, XMM$-$Newton is one of the rare instruments that is capable of observing a particular source in multiple bands 
simultaneously, in this case in the Optical, UV and X-ray bands. The X-ray collecting area of the telescope is huge 
and targets can be exposed for very long periods without any interruption, allowing this satellite to produce sensitive 
observations in the energy range from 0.15 to 15 keV. The EPIC cameras provide  imaging observations taken with a very 
large (30 arcmin) field of view (FOV). These specifications make this satellite perfect for our purpose of studying the 
relation between different bands and their  emitting regions. XMM$-$Newton has observed 3C 273 for over a decade in time 
in X-ray, optical and UV making it possible for us to study variability on different time scales. 

In this paper we have used XMM$-$Newton satellite data to study the behaviour of the source in all three bands. 
Observation data files (ODFs) were downloaded from the on-line XMM-Newton science archive. For reduction of these data 
we have used XMM-Newton Science Analysis Software (SAS) version 12.0.1. The on board system of XMM$-$Newton contains the
European Photon Imaging Camera (EPIC), which consists of two CCD arrays, MOS and pn. Here we have used EPIC/pn camera 
observations for our purpose, since the sensitivity of EPIC/pn is much higher than that of the MOS camera (e.g. 
Str{\"u}der et al.\ 2001). XMM--Newton has data for the source 3C 273 from June 13, 2000 to July 16, 2012 which is 
comprised of a total of 34 observations in XMM--Newton science archive. However, out of these 34 observations two observation 
IDs do not contain any data, and pn data was not available for four other observation IDs. Three more observations were 
excluded because of their bad image quality. So here we have studied 25 observations for the source 3C 273. 

\subsection{Optical/UV Data}

The Optical/UV Monitor Telescope (OM hereafter) on board  XMM--Newton provides the facility to observe simultaneously 
a particular source in optical and UV bands along with the X-ray bands, with a very high imaging sensitivity 
(e.g. Mason et al.\ 2001). This makes possible the study of the relationship between the emission mechanisms corresponding 
to these bands. The OM can collect data with time resolution of 0.5 s for the wavelength range 170 nm -- 650 nm with six 
broad band filters, three in the optical and three in the UV, with a FOV covering  17 arcmin of the central region of the 
X-ray telescope FOV. The optical {\it U, B, V}, filters collect data in the wavelength ranges 300--390 nm, 390--490 nm, 
510--580 nm, respectively and the ultraviolet {\it UVW2, UVM2, UVW1} filters collect data in the wavelength ranges 180--225 
nm, 205--245 nm, and 245--320 nm, respectively.  3C 273 was observed with OM for differing numbers of exposures for different 
observing periods  (see Table 1 for details). The source was simultaneously observed with both X-ray and Optical/UV telescopes 
for a total of 16 observation IDs.  

\begin{table*}

{\bf Table 1. Summary of XMM-Newton data for 3C 273}
\small

\begin{tabular}{lcccccclcl} \hline \hline
 Date of Obs.  & Obs.ID     &Revolution &Window     & GTI    &Pile up &Filter     & $\mu$            &OM filter$^1$&Exposures$^2$    \\
 dd.mm.yyyy    &            &           &Mode       &(ks)    &        &           &(counts $s^{-1}$) &             &              \\\hline
 13.06.2000    & 0126700101 & 94        &Full Frame & Nil   &Nil    & MEDIUM    & Nil             & 6           & 5        \\
 13.06.2000    & 0126700201 & 94        &Full Frame & Nil   &Nil    & MEDIUM    & Nil             & 6           & 5 \\
 13.06.2000    & 0126700301 & 94        &Small      & 64.9   & No     & MEDIUM    & 47.04$\pm$0.81   & 1,2,3,4,5,6 & 29    \\
 15.06.2000    & 0126700401 & 95        &Full Frame & Nil   &Nil    & MEDIUM    & Nil             & 6           & 5  \\
 16.06.2000    & 0126700501 & 95        &Full Frame & Nil   &Nil    & MEDIUM    & Nil             & 6           & 5   \\
 15.06.2000    & 0126700601 & 95        &Small      & 29.6   & No     & MEDIUM    & 45.41$\pm$0.80   & 4,5,6       & 15    \\
 15.06.2000    & 0126700701 & 95        &Small      & 29.9   & No     & MEDIUM    & 44.16$\pm$0.80   & 1,2,3       & 15   \\
 17.06.2000    & 0126700801 & 96        &Small      & 60.1   & No     & MEDIUM    & 44.26$\pm$0.79   & 1,2,3,4,5,6 & 30    \\
 13.06.2001    & 0136550101 & 277       &Small      & 88.5   & Yes    & MEDIUM    & 21.58$\pm$0.58   & 1,2,3,4,5,6 & 12   \\
 16.12.2001    & 0112770101 & 370       &Small      &  4.9   & No     & THIN1     & 71.95$\pm$1.00   & Nil        & Nil    \\
 22.12.2001    & 0112770201 & 373       &Small      &  4.9   & No     & THIN1     & 69.50$\pm$0.99   & Nil        & Nil      \\
 09.01.2002    & 0137551001 & 382       &Burst      & Nil   & Nil   & MEDIUM    & Nil             & 6           & 1    \\
 07.07.2002    & 0112770601 & 472       &Small      &  4.9   & No     & THIN1     & 53.93$\pm$0.87   & Nil        & Nil     \\
 17.12.2002    & 0112770801 & 554       &Small      &  4.9   & No     & THIN1     & 77.54$\pm$1.04   & Nil        & Nil     \\
 05.01.2003    & 0112770701 & 563       &Small      &  4.9   & No     & THIN1     & 65.32$\pm$0.96   & Nil        & Nil     \\
 05.01.2003    & 0136550501 & 563       &Small      &  8.4   & No     & MEDIUM    & 63.22$\pm$0.94   & 2,3         & 2    \\
 18.06.2003    & 0112771001 & 645       &Small      &  5.4   & No     & THIN1     & 80.24$\pm$1.09   & Nil        & Nil   \\
 07.07.2003    & 0159960101 & 655       &Small      & 58.0   & Yes    & THIN1     & 26.05$\pm$0.57   & 1,3,4,5,6   & 5 \\
 08.07.2003    & 0112770501 & 655       &Small      &  8.0   & No     & THIN1     & 70.66$\pm$1.04   & Nil        & Nil    \\
 14.12.2003    & 0112771101 & 735       &Small      &  8.3   & No     & THIN1     & 53.34$\pm$0.87   & Nil        & Nil    \\
 30.06.2004    & 0136550801 & 835       &Small      & 19.7   & No     & MEDIUM    & 45.27$\pm$0.81   & 1,2,3,4,5,6 & 6  \\
 10.07.2005    & 0136551001 & 1023      &Small      & 27.5   & No     & MEDIUM    & 49.68$\pm$0.84   & 1,2,3,4,5,6 & 6    \\
 12.01.2007    & 0414190101 & 1299      &Small      & 76.5   & Yes    & MEDIUM    & 21.42$\pm$0.52   & 1,2,3,4,5,6 & 9    \\
 25.06.2007    & 0414190301 & 1381      &Small      & 31.9   & No     & MEDIUM    & 45.90$\pm$0.80   & 1,2,3,4,5,6 & 8    \\
 08.12.2007    & 0414190401 & 1465      &Small      & 35.3   & Yes    & MEDIUM    & 38.16$\pm$0.69   & 1,2,3,4,5,6 & 6    \\
 09.12.2008    & 0414190501 & 1649      &Small      & 40.4   & Yes    & THIN1     & 21.57$\pm$0.52   & 1,2,3,4,5,6 & 6   \\
 20.12.2009    & 0414190601 & 1837      &Small      & 31.3   & Yes    & THIN1     & 24.36$\pm$0.57   & Nil        & Nil   \\
 10.12.2010    & 0414190701 & 2015      &Small      & 35.8   & Yes    & THIN1     & 18.29$\pm$0.48   & 1,2,3,4,5,6 & 6  \\
 12.12.2011    & 0414190801 & 2199      &Small      & 42.8   & No     & THIN1     & 47.48$\pm$0.82   & 1,2,3,4,5,6 & 6   \\
 16.07.2012    & 0414191001 & 2308      &Small      & 25.5   & No     & THIN1     & 41.22$\pm$0.93   & 4,5,6       & 3  \\\hline

 \end{tabular}     \\
 $\mu$ = mean count rate; \\
 GTI = Good Time Interval  \\
 $^1$ 1 = {\it UVW2}, 2 = {\it UVM2}, 3 = {\it UVW1}, 4 = {\it U}, 5 ={\it  B}, 6 = {\it V}    \\
 $^2$ Total number of exposures combining all six filters \\
\end{table*}

\section{Data Analysis Techniques}

\subsection{Light Curve generation}

The EPIC/pn camera takes images of a target source for the energy range 0.15$-$15 keV, but data above 10 keV
are dominated by strong proton flaring. X-rays emitted below 0.3 keV  are highly affected by 
hydrogen in the direction of 3C 273 in our own galaxy. Moreover, the on-axis effective area of the EPIC pn 
camera play an important role in the performance of the X-ray mirrors. These mirrors can reflect X-ray photons most
efficiently in the energy range from 0.1 to 10 keV. Therefore, for better quality and higher accuracy of the data set,
we  choose the energy range from 0.3$-$10 keV for the light curve extraction. By doing this we automatically
eliminate the effect of absorption in the lower energy range and the background noise in the higher energy range. All EPIC/pn 
observations we have used were taken in small window (SW) mode.
 
Event files of the pn detector were generated through {\it epchain}. Before generating a cleaned event list, first 
we generated a light curve for energy range of 10--12 keV and saw the soft proton flares effect on this and then 
generated a good time interval file using the {\it TABGTIGEN} tool which contains information of the good times that are 
free from soft proton flares. In the next step, we used the event list file and GTI files as input to obtain cleaned event 
files and we later filtered the data using the condition $(PATTERN \leq 4)$ energy range 0.3$-$10 keV and $(FLAG $=$ 0)$. 
With cleaned event files, we have generated source images and for source event files, the bright circular source 
region of 40 arcsec radius is selected in such a way that 90 per cent of the source flux lies inside the circle, for 
background event files, we selected a circular region of maximum possible size (which ranged between 
40$-$50 arcsec for different images/CCD chips), but as far away as possible on the same CCD chip to avoid any contribution 
from the source region.

Pile up is a common problem for very bright sources like our blazar and was an issue for seven observation IDs. 
Pile up occurs when two or more X-ray photons fall on a single 
CCD pixel before the instrument's read out time, and as a result, the CCD will recognize these  separate photons as a 
single one with energy equal to the sum of the individual energies of the two or more photons. It also may arise when more than 
one photon is detected by two or more adjacent pixels during a single read-out cycle. This pile up problem is detected 
by using {\it epatplot} routine for all the observation IDs and later they were removed by selecting an central annulus 
region instead of circular region for both the source events file and background events file as  used before in non-pile 
up affected observations on the same CCD chip. Since all observations we have used here in our study were carried 
out in small window mode (time resolution $=$ 5.7 ms), we used a binning of 114 seconds, which is an integer multiple of the 
instrument time resolution.  We obtain the final corrected events list by subtracting the background counts from the source 
counts. The events list files obtained by using the {\it epiclccorr}
task contains some null points occurring when the window of the CCD chip was being closed during that exposure time. 
Also there are some fictitious points which show significantly larger error bars than the others with a peculiar 
deviation in counts from the normal order at the start or end in the events list; these originateswhen the source 
exposure time is less than 70 per cent of the window flag time. These points were removed to generate the final corrected 
lists. After correcting all these issues, we finally plot the light curves for observations. The 3C 273 public data 
over the 12 year span is given in Table 1.

We take the imaging mode data of OM for our purpose, which we later reprocessed with the perl script {\it omichain}, 
which starts different tasks within SAS for full processing of OM data. After running the metatask {\it omichain}, we get a 
observation source list which contains the calibrated data with their errors. The source list contains several objects 
and their corresponding data points are treated as background points.  The objects near to the source co-ordinate appear in 
the source FOV with similar RA and DEC values. We identify our source from this object list by selecting the 
right RA and DEC value for our source. 

Since the blazar is among the brightest objects in the OM FOV, the source will have very large counts rate in all filters. 

\begin{figure*}

\epsfig{figure=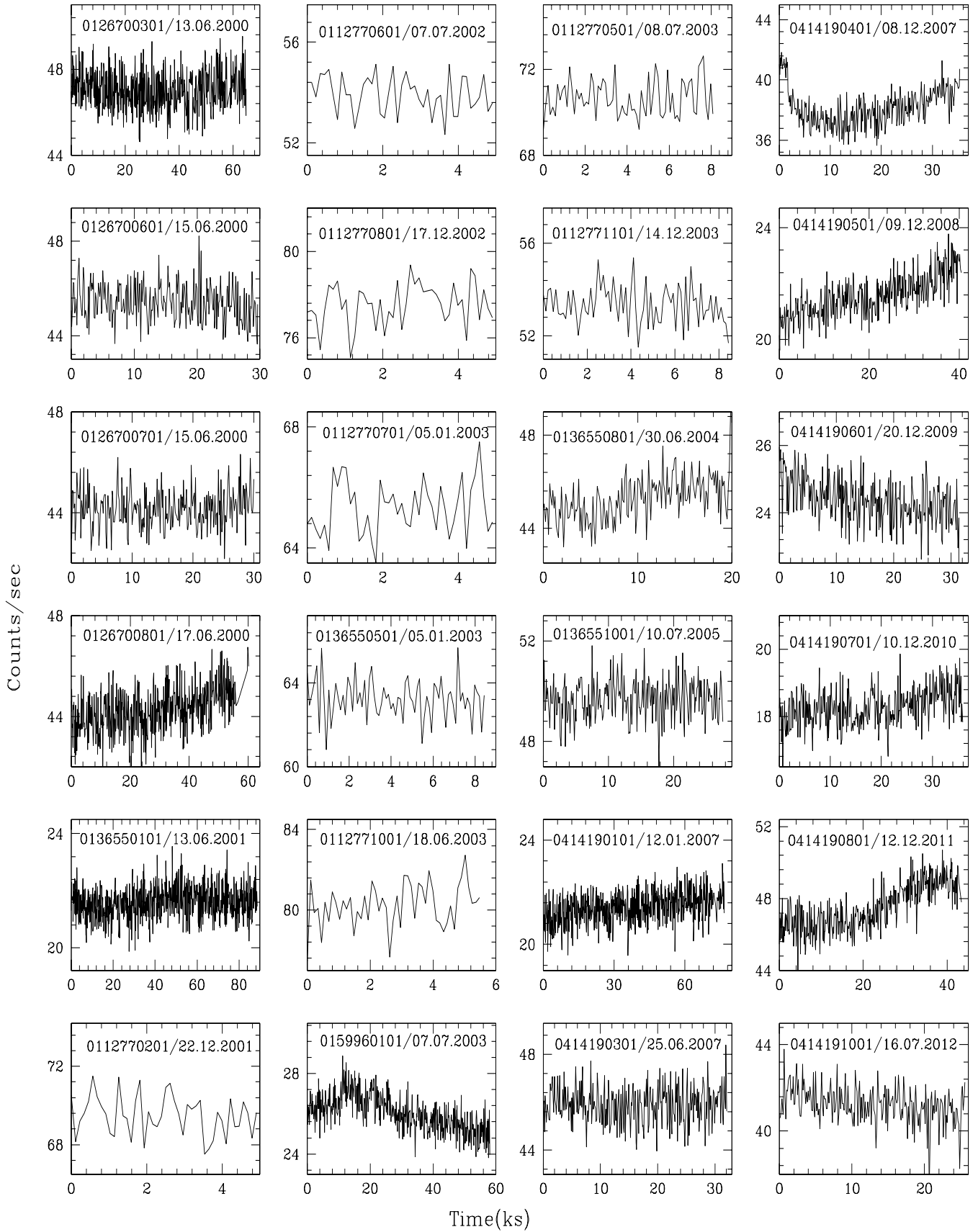, height=9.0in,width=7.5in,angle=0}
\caption{The variation of X-ray photons with time for 3C 273,  for observations carried out at different times. The caption on each light curve represents observation ID and observing date, respectively. These light curves were extracted using 114 seconds time binning in the energy range, 0.3$-$10 keV.}
\label{fig:1}
\end{figure*}

\subsection{Excess Variance}

Blazars show very rapid flux variability in all wave bands from radio to $\gamma-$rays. 
To measure the strength of variability, the commonly used parameters are excess variance, $\sigma_{XS}^2$ and fractional 
rms variability amplitude, $F_{var}$,
(e.g.\ Edelson et al.\ 2002). Each light curve obtained from different exposures contains information on flux measurements 
(count rates) with its uncertainties caused by measurement errors for different time intervals. Excess variance gives 
the intrinsic variance of the source by removing the variance due to measurement errors in each individual flux measurement. 
When the light curve consist of total number $n$ of flux measurements $x_{i}$,  at times $t_{i}$, with corresponding errors 
in measurements $\sigma_{err,i}$, then the excess variance is calculated as

\begin{equation}
\sigma_{XS}^{2} = S^{2} - \overline{\sigma_{err}^2},
\end{equation}
where $\overline{\sigma_{err}^{2}}$ is the mean square error and $S^2$ is the sample variance of the light curve given by 

\begin{equation}
S^2 = \frac{1}{n-1}\sum_{i=1}^{n} (x_{i}-\overline{x})^2
\end{equation}
and
\begin{equation}
\overline{\sigma_{err}^2} = \frac{1}{n} \sum _{i=1}^{n} \sigma_{err,i}^{2}.
\end{equation}

The fractional rms variability gives the average variability amplitude with respect to the mean flux of a source,
which is nothing but the square root of the normalized excess variance, $\sigma_{NXS}^{2}$, which is 

\begin{equation}
\sigma_{NXS}^{2} = { \frac {\overline{\sigma_{XS}^2}}  {\overline{x}^2}}
\end{equation}
so
\begin{equation}
F_{var} = \sqrt{ \frac{ S^{2} -
\overline{\sigma_{err}^{2}}}{\bar{x}^{2}}}
\end{equation}
The uncertainty on $F_{var}$ has been calculated using Monte Carlo method for $n$ total  events 
(e.g. Vaughan et al. 2003) and is  given by

\begin{equation}
(F_{var})_{err}=\sqrt{ \left\{ \sqrt{\frac{1}{2n}} \frac{ \overline{\sigma_{err}^{2}}
}{\bar{x}^{2}F_{var} } \right\}^{2}+\left\{ \sqrt{\frac{\overline{\sigma_{err}^{2}}}{n}}\frac{1}{\bar{x}}\right\}^{2}}.
\end{equation}

\section{Results}

\subsection{Intraday Variability in X-ray light curves}

We have searched for IDV in X-ray light curves for those observations given in Table 1 that provided
good data
and those 24 light curves are plotted in Fig.\ 1. In Table 2, we have reported the IDV variability parameters
which are calculated using excesses variance. It can be clearly seen that the IDV amplitude in the 
individual observations are low, usually less than 1 per cent and often consistent with 0, given the 
errors on $F_{var}$. The highest variability among all the observations taken during 
this time span is seen in the observation ID 0159960101 in 2003, and is 2.60 per cent, though the variabilities
are essentially as large during two other observations (0414190401 and 0414190501).  We note that
the longer GTI observations produce higher probabilities of detecting clear variations.

In blazars, it is believed that the  X-ray emission is a result of synchrotron radiation from highly 
energetic electrons moving with relativistic speeds in the jet.  It can either be direct synchrotron
radiation or results from inverse Compton scattering of lower energy photons off the synchrotron
emitting electrons. In case of FSRQs X-ray emissions mainly come from inverse-Compton emission, 
but here in 3C 273, which is a acceptional case regarding the presence of both Seyfert like and blazar like 
components in its X-ray spectra, X-rays below 2 keV is due to coronal emission from the accretion disk and 
above 2 keV is dominated by contribution from jet emission (e.g.\ Page et al.\ 2004; Pietrini \& Torricelli-Ciamponi 
2008; Haardt et al.\ 1998). Blazars show complex emission behaviour in X-ray flaring states. During these states 
the object can show both soft and hard lags depending on the time required for cooling and acceleration by radiating 
electrons. From the smallest variability timescale in X-ray band, the size of the central emitting regions of the 
blazar can be estimated. If IDV is found during the quiet state of the blazar, when the jet is weak or absent, 
then it may be attributed as result of changes occurring in the accretion disk, which can constrain the 
structure of the central engine on linear scales (Gupta et al.\ 2009). Intra-day X-ray variability may also 
be due to instabilities on or above the accretion disk, caused by its magnetic field that can also lead to 
generation of oscillations in the disk. 

\begin{table}
{\bf Table 2. X$-$ray variability parameters} 
\small

\begin{tabular}{lccc} \hline \hline

Obs.IDs         &Variance    &$\sigma_{XS}^2$ &F$_{var}$ (\%)    \\
                &            &                &    \\\hline  
0112770101	& 0.82	     & 0.19	& 0.61$\pm$0.40     \\
0112770201      & 1.04	     & 0.06	& 0.34$\pm$0.66	   \\
0112770501	& 0.71	     & 0.37	& 0.86$\pm$0.27    \\

0112770601	& 0.56	     & 0.20	& 0.83$\pm$0.42	    \\
0112770701	& 0.79	     & 0.13	& 0.54$\pm$0.47	    \\
0112770801	& 0.95	     & 0.13	& 0.47$\pm$0.46    \\
0112771001	& 1.16	     & 0.02	& 0.19$\pm$1.00    \\
0112771101	& 0.69	     & 0.08	& 0.51$\pm$0.46    \\
0126700301	& 0.76	     & 0.11	& 0.69$\pm$0.15   \\
0126700601	& 0.63	     & 0.01	& 0.16$\pm$0.85   \\
0126700701	& 0.55       & 0.09	& 0.67$\pm$0.24	  \\
0126700801	& 0.77       & 0.15	& 0.88$\pm$0.14    \\
0136550101	& 0.30	     & 0.03	& 0.80$\pm$0.21	  \\
0136550501	& 0.90	     & 0.01	& 0.12$\pm$1.47    \\
0136550801	& 0.85	     & 0.20	& 0.99$\pm$0.22    \\
0136551001	& 0.60	     & 0.10     & 0.65$\pm$0.23   \\
0159960101	& 0.79	     & 0.46	& 2.60$\pm$0.11	  \\
0414190101	& 0.33	     & 0.06	& 1.15$\pm$0.17   \\
0414190301	& 0.62	     & 0.03	& 0.36$\pm$0.37    \\
0414190401      & 1.44	     & 0.96	& 2.57$\pm$0.12   \\
0414190501	& 0.56	     & 0.28	& 2.47$\pm$0.16	  \\
0414190601	& 0.48	     & 0.15	& 1.61$\pm$0.20   \\
0414190701	& 0.30	     & 0.07	& 1.48$\pm$0.23   \\
0414190801	& 1.58	     & 0.91	& 2.01$\pm$0.10	   \\
0414191001	& 0.69	     & 0.18	& 1.02$\pm$0.28    \\\hline

\end{tabular}     \\

\end{table}

\subsection{Long term variation in Optical/UV \& X-ray bands}

The source has been observed for more than a decade, which makes it possible for us to study its 
long term variability behaviour in X-rays as well as in Optical/UV bands.  We examine long term behaviour 
of the source 3C 273 by combining the mean count rates for each observation. The temporal 
behaviour of the source in between 2000 and 2012, in 2 X-ray and 6 Optical/UV bands are plotted
in different panels of Fig.\ 2.  By visual inspection, variability is clearly seen in all of these bands.  

We split the X-ray light curve into two light curves with the energy ranges being 0.3--2.0 keV for soft X--rays 
and 4.0--10 keV for hard X--rays. Previous studies (Page et al. 2004, Pietrini and Torricelli-Ciamponi, 2008, Haardt 
et al.\ 1998) have found that the disk emission is the origin of X-ray emission below 2 keV in 3C 273, where disk photons 
scattered off the hot corona above the disk, while emission above 2 keV comes from inverse-Compton scattering of seed 
photons by the relativistic jet. So we consider the energy range for soft X--ray band is 0.3--2 keV to be dominated
by coronal emission but for the  main contribution from the jet, we consider the hard band energy range 4--10 keV, 
in order to reduce contamination from the X--ray corona. Both the soft and hard X--ray light curves show very similar 
variability patterns. It seems that there was a small flaring in both the X-ray bands in June 2007 with almost no time lag. 
A similar kind of flare is seen in December 2007, in all the optical/UV bands when the X-ray fluxes are lower. 
Though we clearly do not have enough data points to formally  calculate any time lag it is not unreasonable to 
estimate a time delay of  the flare in Optical/UV bands with respect to the X-ray bands by visual inspection 
which give the delay of about six months. However, we cannot claim  whether there was a correlation between 
these two emission peaks or if they originate from two independent emitting regions. We did not find
any strong flares. But there are evidences of flares from the source on different
occasions. Fan et al.\ (2014) studied the photometric observation of 3C 273 in optical bands from 1998 to 2008, 
where they reported detection of two peaks in February 2006 and March 2008 in R band. Beaklini et al.\ (2014) 
reported a flare at 7 mm wavelength with Itapetinga Radiotelescope, in Brazil in 2010, which they claim as counterpart 
of gamma-ray flare detected in September, 2009 by ther Fermi / Large Area Telescope (LAT)with a delay of 170 days. 
Frequent flares were detected by Fermi/LAT  in gamma-ray bands in the period September 2009 to April 2010 (Rani et al.\ 2013).

Over the long term, 3C 273 showed large RMS variability amplitudes. The RMS variability amplitude for 
{\it UVW2, UVM2, UVW1, U, B, V} bands and soft and hard X-ray bands are, in per cent $\sim$ 76, 76, 68, 68, 68, 
35, 42, and 36, respectively, with respective perccentage errors in variability amplitude of  0.016, 0.0104, 0.0867, 
0.76, 0.30, 0.086, 0.475 and 0.983. Thus we conclude that the source is highly variable in both the optical/UV and 
X-ray domains during the above mentioned time period. 

The overall change in the total X-ray flux with time during this observing period of 3C 273 is shown in Fig.\ 3. On
visual inspection, if we exclude small portions of the starting and ending period of this observation there is
indication of  a linearly decreasing flux profile with time from December 2001 to December 2010. This data is
fitted with least squares fitting method and the correlation coefficient for the fit calculated with 
MATLAB, is $-0.8076$ and its corresponding null-hypothesis is 0.001. In  Fig.\ 3, the excluded parts of 
these observations are separated by vertical dotted lines.       

\begin{figure*}
\centering
\includegraphics[width=7.5in,height=6.5in]{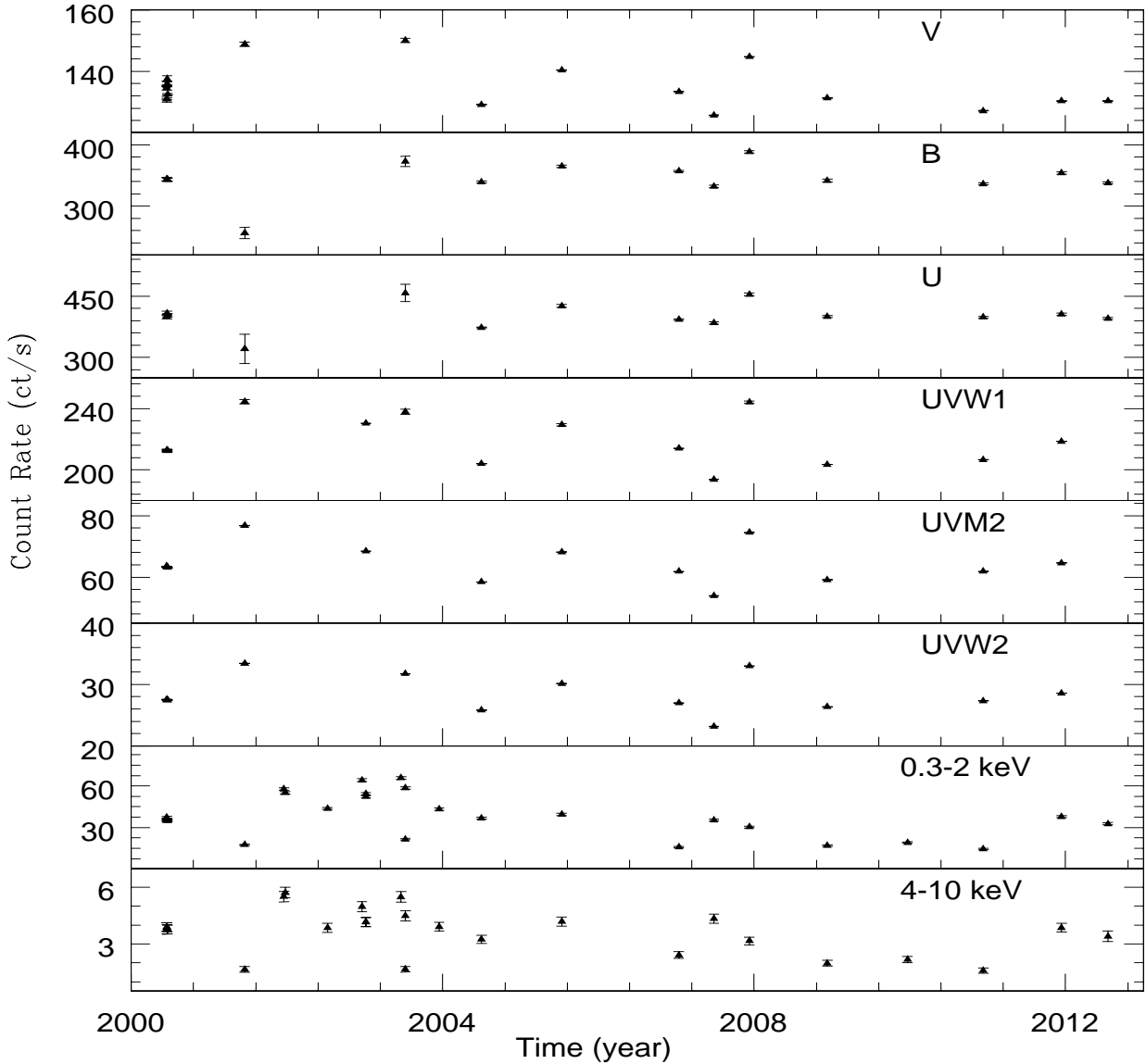}
\caption{Optical/UV and X-ray long term variability (2000-2012). The lower panel shows two X-ray bands, 
hard and soft, respectively from the bottom. Three UV bands with increasing wavelength are plotted just 
above X-ray bands and in the upper panel, optical U, B and V bands  are shown.}
\end{figure*}

\begin{figure}
\psfig{figure = 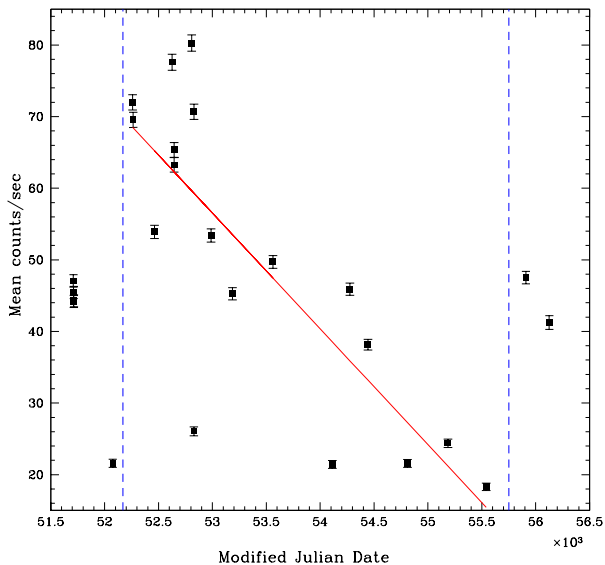 ,scale=0.4}
\caption{Long term X-ray variability  of 3C 273 (2000-2012)}
\label{fig7}
\vspace*{.7cm}
\end{figure}

\subsection{Possible connections between Optical/UV and X-ray bands}

From Fig.\ 2,  visually it appears that the optical and UV bands follow the same variability 
pattern  as do the soft and hard X-ray bands. To further examine  the variability relations between these 
optical, UV and X-ray bands, we plotted these bands against each other as shown in Fig.\ 4. 
The upper left panel of Fig.\ 4 explicitly shows that soft and hard X-ray bands are well correlated. In the same 
manner, to check the correlation in flux variation within broad bands, we plot the optical V band against 
the B band and in similar way plot the  UWV2 band against the UVW1 (shortest wavelength UV filter for the OM telescope). 
Clearly good correlations are found  for all  three intra-bands cases. 

Similarly, Optical/UV and X-ray flux variabilities
are checked by plotting soft X-ray vs V, and soft X-ray vs UVW1 which are shown in the left and middle 
panels of the bottom row of Fig.\ 4. These plots do not show any correlations, but there is  good
correlation between the V and UVW1 counts. From this analysis, we 
conclude that the variability within the X-ray, optical \& UV  bands are closed related to each other and  further that
the optical emission follows the same trend as UV emission. But there is no obvious correlation  between the variability 
trends for the X-ray and the Optical/UV emissions. We also calculated correlation coefficients; they and their 
corresponding null hypothesis tests for each pair 
of combinations are given in Table 3.

\begin{table}
{\bf Table 3. Variability Correlations between Optical, UV and X-ray bands} 
\small

\begin{tabular}{lcc} \hline \hline

Bands                 &Correlation coefficient  &Null hypothesis   \\
                      &                            &                  \\\hline  
Soft$-$Hard X-ray     & 0.9207	                   &0.0 	      \\
B-V	              & 0.8845                     &0.0001               \\
UVW1$-$UVW2           & 0.9862                     &0.00          \\
Soft X-ray$-$V        &-0.2308                     &0.4272             \\
Soft X-ray$-$UVW1     &-0.0012                     &0.9968            \\
UVW1$-$V              & 0.9376                     &0.0               \\\hline

\end{tabular}     \\
\end{table}

\begin{figure*}
  \centering
  \subfloat[][]{\includegraphics[width=0.32\textwidth]{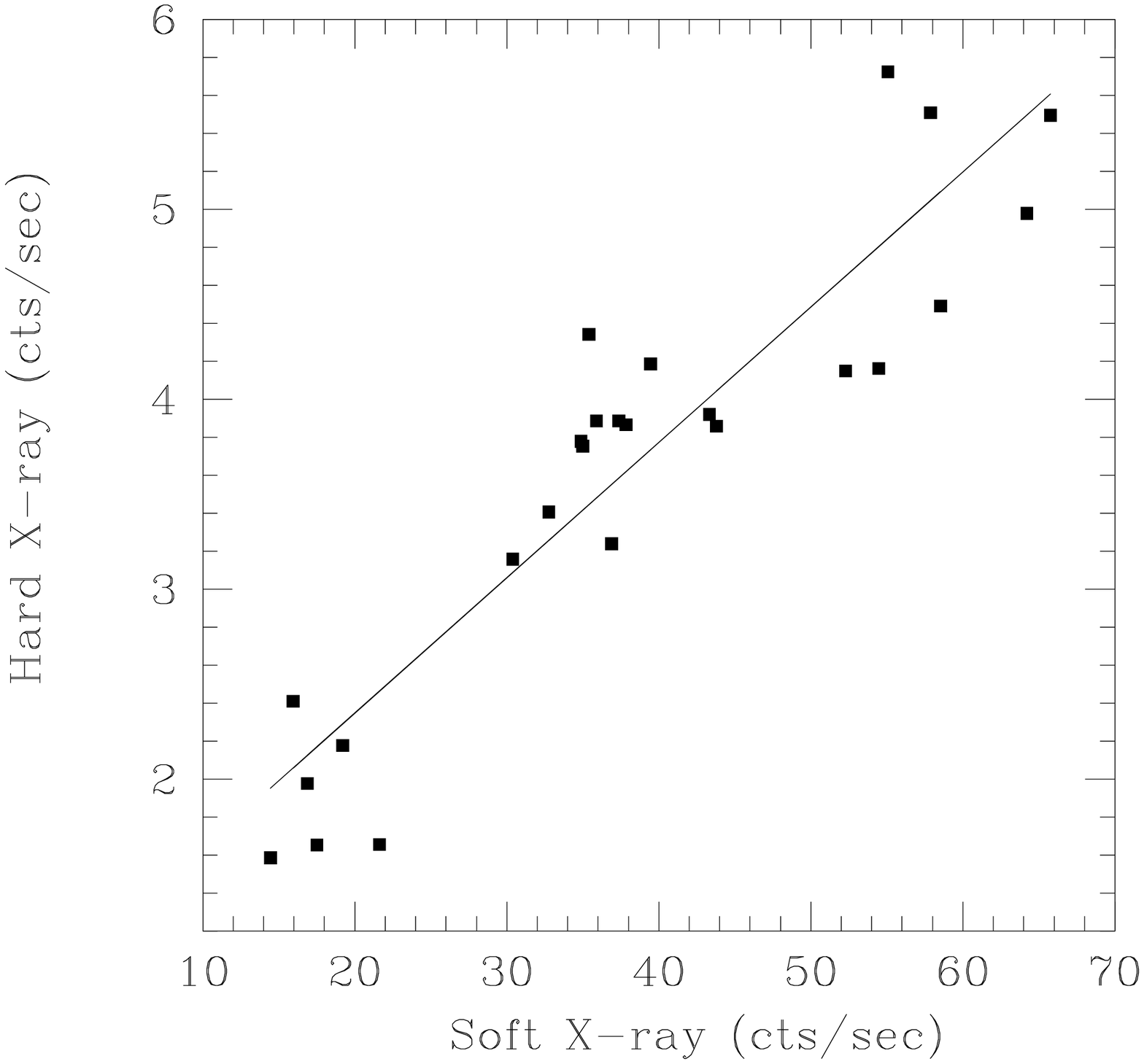}}\quad
  \subfloat[][]{\includegraphics[width=0.32\textwidth]{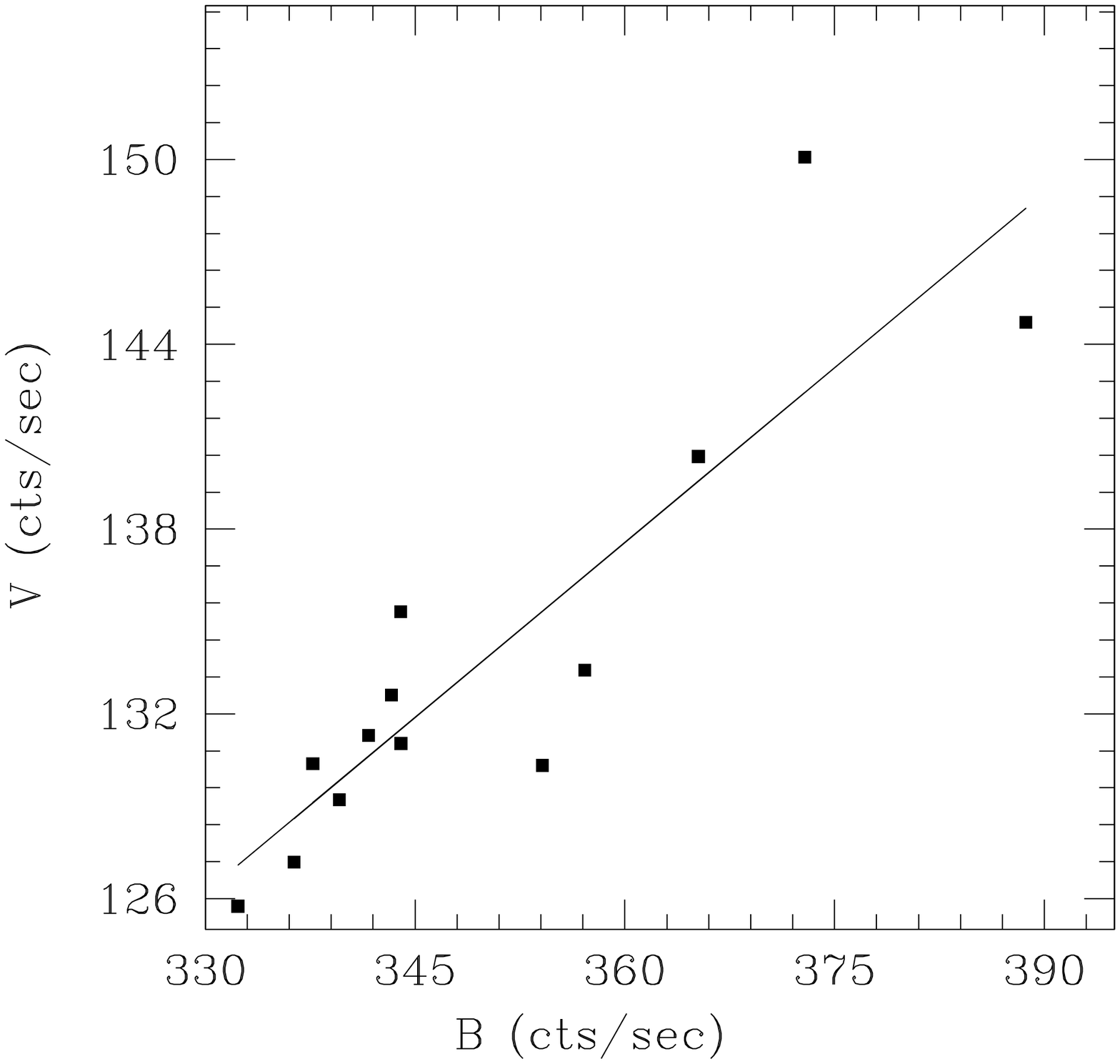}}\quad
  \subfloat[][]{\includegraphics[width=0.32\textwidth]{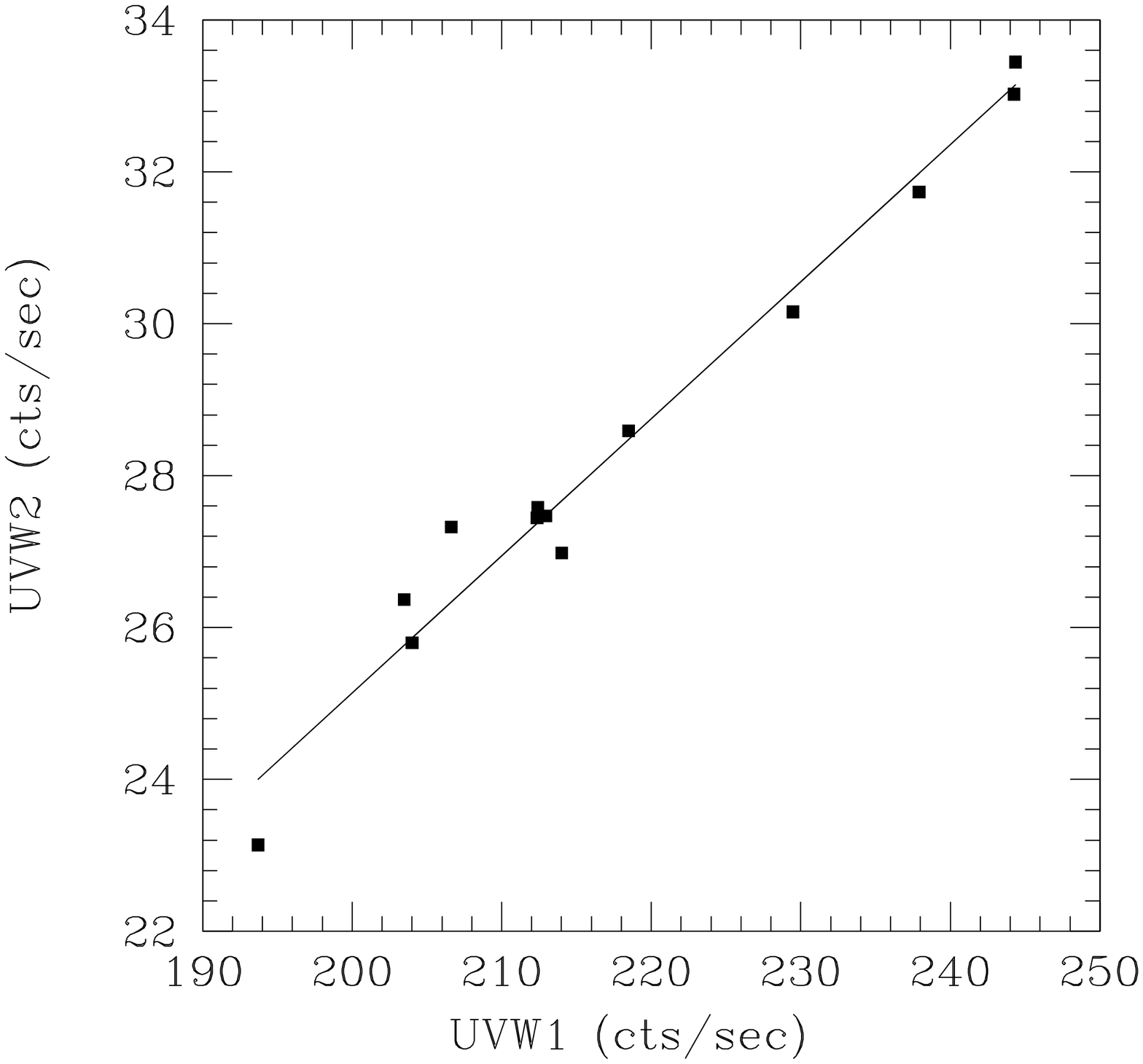}}\\
  \subfloat[][]{\includegraphics[width=0.32\textwidth]{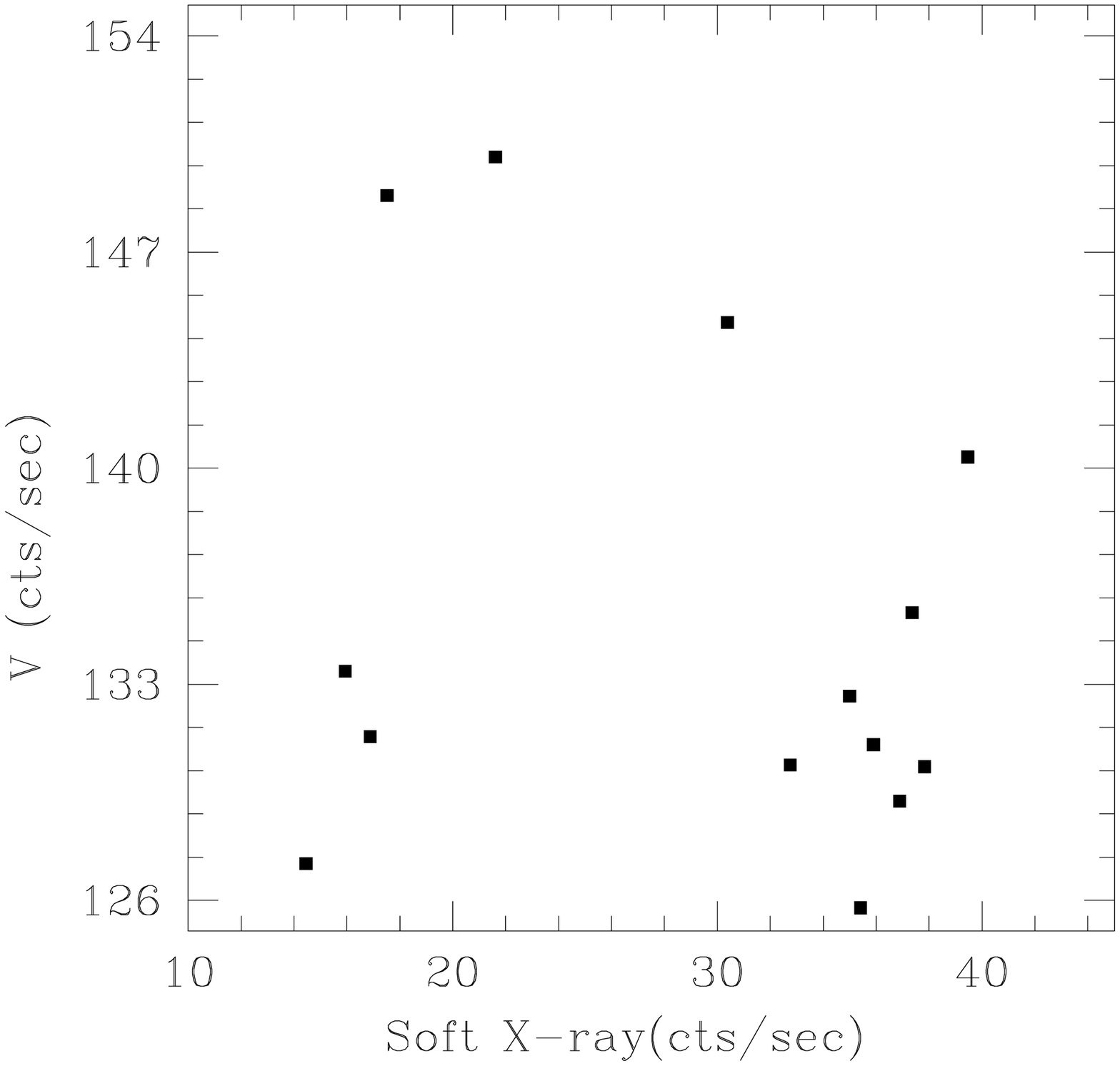}}\quad
  \subfloat[][]{\includegraphics[width=0.32\textwidth]{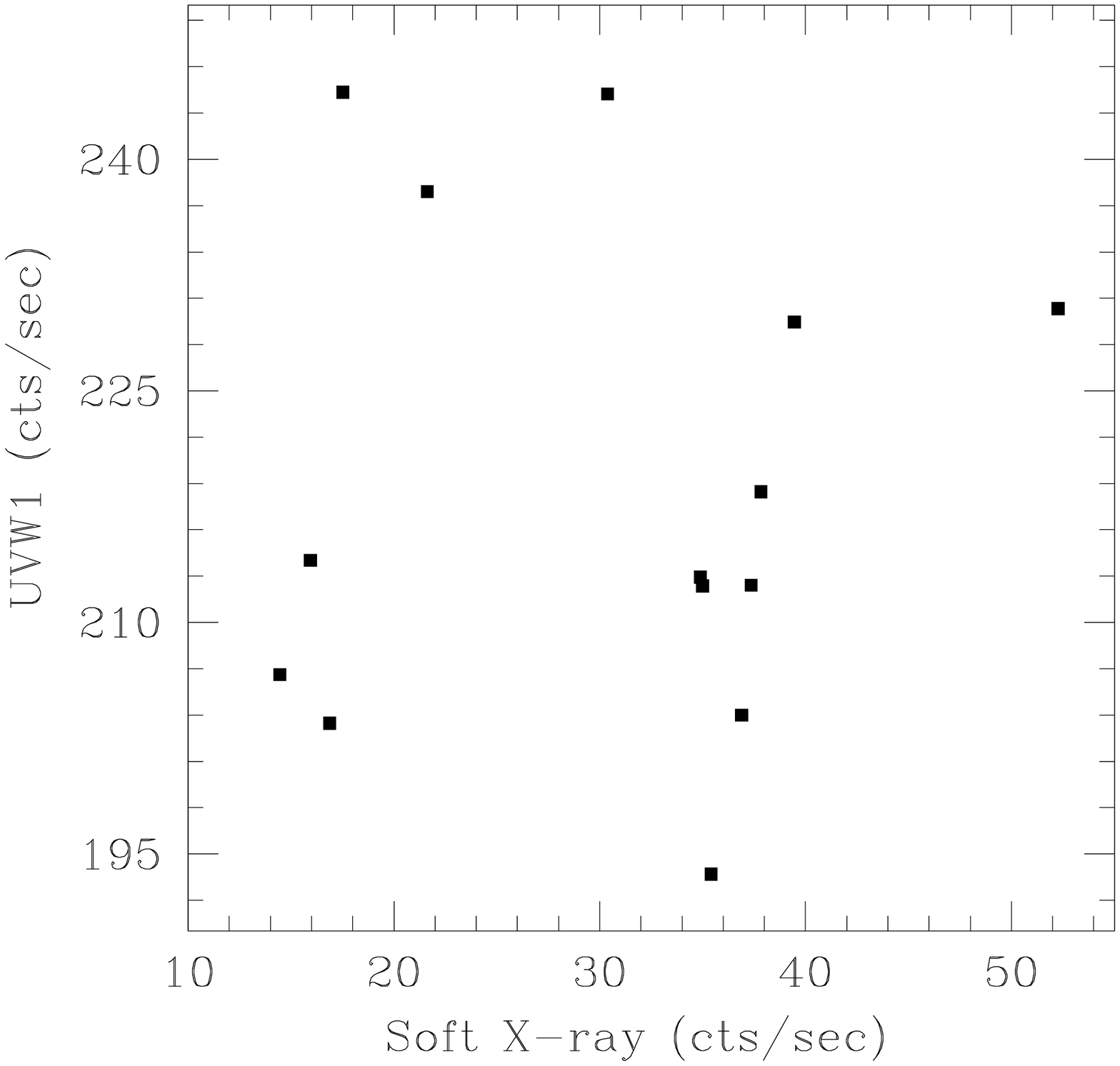}}\quad
  \subfloat[][]{\includegraphics[width=0.32\textwidth]{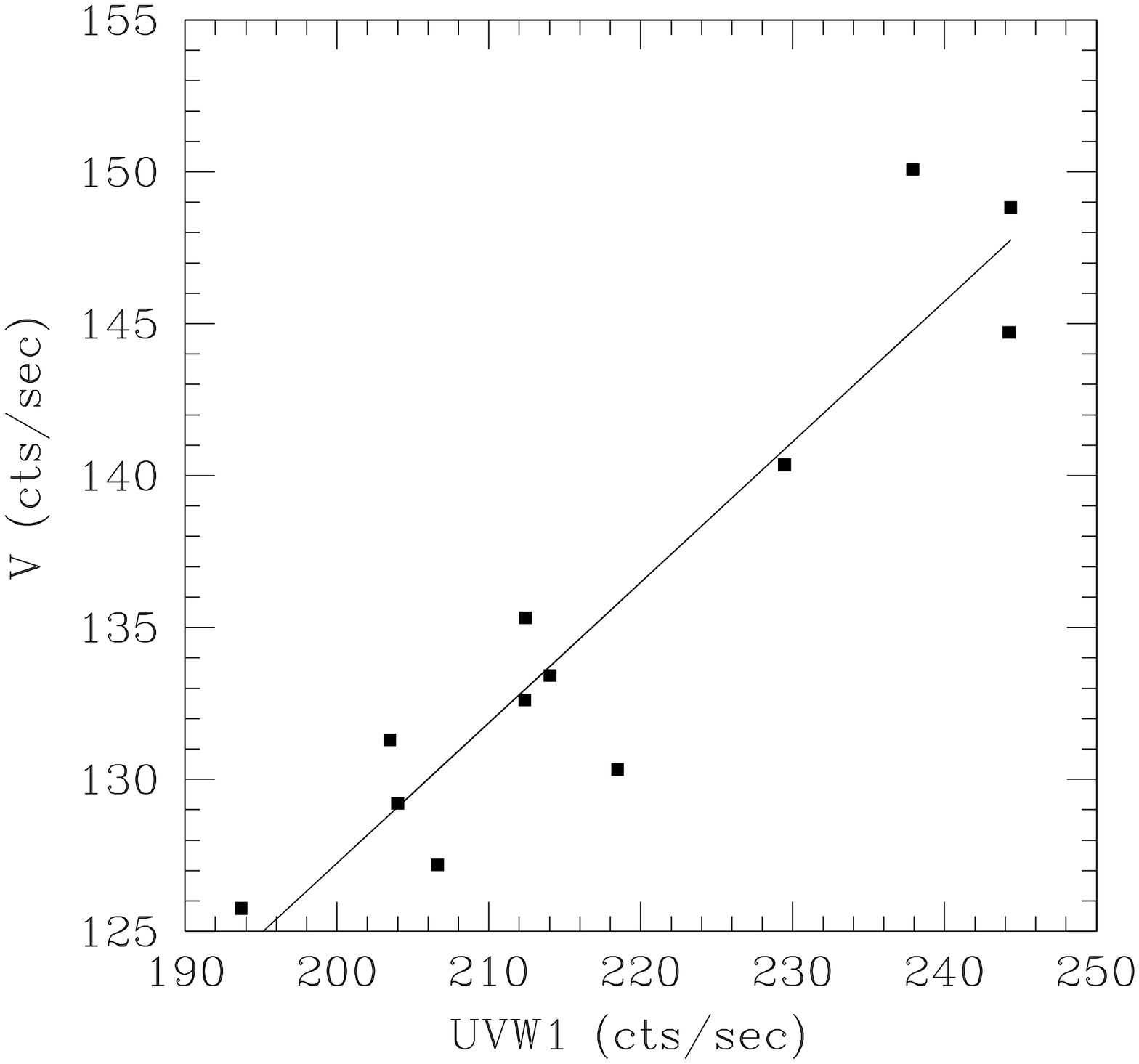}}
  \caption{Correlations in variability between different optical, UV and X-ray bands.}
  \label{fig:sub1}
\end{figure*}

\subsection{X-ray Hardness Ratio}

The hardness ratio 
is evaluated by dividing hard band (4$-$10 keV) count rates by soft band (0.3$-$2 keV) count
rates. To study the spectral evolution of the source, we examined  patterns in the hardness ratio as a 
function of 0.3$-$10 keV count rates in the X-ray domain for different time spans. 
During this period of 12 years for which XMM$-$Newton observed 3C 273, the source has 
continuously evolved through spectral variations that are plotted 
in Fig.\ 5. 
These are best understood in terms of changes in the relative strength of
particle acceleration and synchrotron cooling processes in X-ray 
emitting regions in 3C 273. 

The results of this analysis provides a 
model independent spectral variability study of the source.  Any loop occurring in the spectral hardness -- flux 
representation, whether it is clockwise or anti-clockwise gives information about the leading emission
mechanism during that period by considering the corresponding flux emitted at that time. So we plotted the 
hardness ratio points against the total X-ray count rates to see if any any loops appeared in the plot. Once a 
loop was visible in the plot, we terminated the plotting and marked that time span as an epoch and began a search
for subsequent loops. In this
process, we sometimes had to consider a single hardness ratio point as marking both the end of one
loop plot and the  beginning
of another to best indicate  the complete loops.

 The widely accepted picture 
for the X-ray emission is that shock propagation in the relativistic jet is responsible for accelerating particles 
to extremely high energies, and when these highly relativistic particles encounter an inhomogeneous, 
twisted magnetic field in the jet, significant
synchrotron emission occurs into the X-ray band, which dominates the cooling process. In Fig.\ 5, it is seen that in 
Epoch-1, the time span of June 2000 to December 2001, we observed an anti-clockwise loop (or hard-lag). 
This indicates that during that period particles were being accelerated to their  highest speeds. We again have indications
of the particles being accelerated in the internal shock outflow within the jet from epoch-2 (December 2002 to June
2003) and Epoch-3 (July 2003 to January 2007). Excluding Epoch$-$4 (January 2007 to December 2008), 
in Epoch$-$5, which spans over  (6 December 2007$-$16 July 2012) too we found an anticlockwise loop in the spectral$-$flux 
representation indicating dominance of the particle acceleration mechanism. Only Epoch-4 shows a soft-lag and
clockwise loop, which provides evidence of the synchrotron cooling mechanism being dominant during that particular period. 

\begin{figure*}      
\centering
\mbox{\subfloat{\includegraphics[width=3.55in]{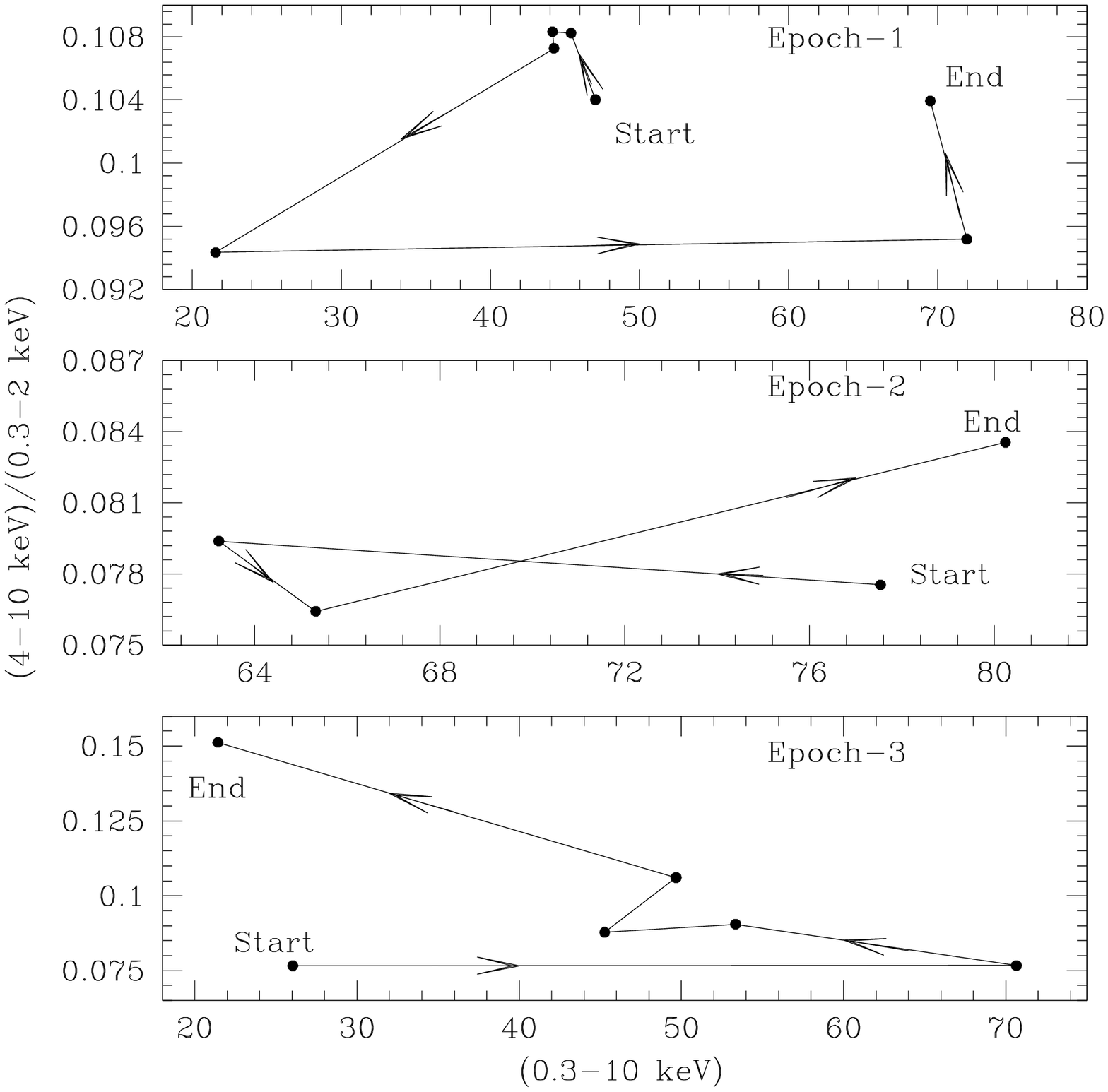}}\quad
\subfloat{\includegraphics[width=3.55in]{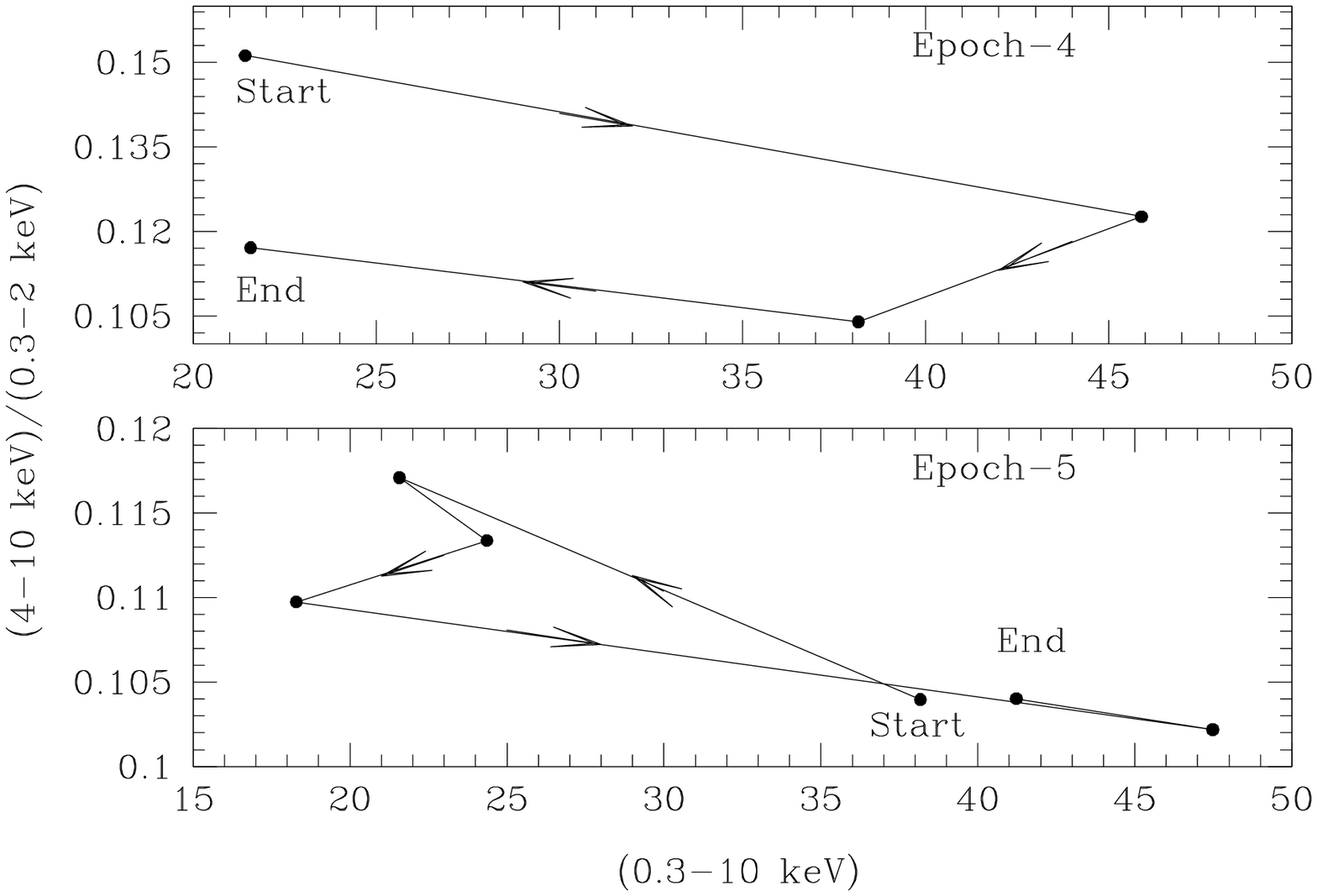} }}
\caption{Spectral evolution of 3C 273 in different epochs. Arrows indicate the directions of the
loops. Here the term epoch in the plots represent different time intervals during which the data were acquired for each corresponding
loop: epoch$-$1 =13.06.2000$-$22.12.2001; epoch$-$2 =17.12.2002$-$18.06.2003; epoch$-$3 =07.07.2003$-$12.01.2007;
epoch$-$4 =12.01.2007$-$ 09.12.2008; epoch$-$5 = 06.12.2007$-$16.07.2012. The clockwise and anticlockwise loops
are distinct and closed or nearly so, presenting a clear evidence of alternate acceleration and cooling mechanism.}
\vspace*{0.7cm}
\end{figure*}

\section{Discussion and Conclusions}

In blazars, it is widely believed that variability is dominated by jet emission, which is significantly 
amplified by relativistic beaming. It has been believed most X-ray variability  originates in a small region 
of the jet  close to the SMBH of the blazar. The propagation of shock waves due to the fluctuations in 
hydromagnetically turbulent plasma in the relativistic jets can explain well the IDV and short term 
variability in blazars (e.g. Marscher 1996, 2014), though some  intra-day variability may be also explained by instabilities 
due to magnetic fields or hot spotS on or above the accretion disk (e.g. Wiita 1996).

We have performed a time series analysis of the extensively studied blazar  3C 273 in X-ray and 
optical/UV bands for the data taken from XMM$-$Newton over a time period of more than a decade. For 
our analysis, we first plotted the $0.3-10$ keV light curves of the source, shown in Fig.\ 1. The 
blazar pointed observation period varied from $4.9$ks for the shortest stare to $88.5$ks for the longest.
 To study the IDV of the source in this period we have estimated the variability amplitude for 
all these observations and gave the results in Table 2.  Only a few 
of these light curves showed even moderate flux variability, which only attained a maximum fractional rms variability 
of 2.6 per cent. Unsurprisingly,   those observations, which were carried out for a longer period of times 
($>$ $25$ ks) gave more detectable variations, with  F$_{var} >$ 1 per cent, while that  lasted  lees than $25$ ks 
showed values of F$_{var}$  below 1 per cent with errors large enough to make them indistinguishable from
null variations.

Over longer times we saw that 3C 273 is highly variable in all bands that XMM$-$Newton measured in
the X-ray, UV and optical. In the Optical/UV bands, the variability amplitude decreases from UV to optical  
and was minimum for the V band, which is the lowest frequency measured.  
This can be understood simply in terms of the expectation that  for synchrotron emission the more
highly energetic particles  cool more rapidly via the synchrotron and inverse Compton processes than do the less
energetic ones.  However, in the case of X-ray emission, variability in the soft band is 
higher than that in the hard band. 

From our study of the variability relations between different bands, we see that soft and hard X-ray bands follow 
each other tightly,  as shown in Fig.\ 4(a), indicating that their emission originates 
from same electron population.  This doesn't answer the question as to whether the  soft band follows the hard band 
of vice versa.  The answer could in principle be found  by computing the discrete correlation function 
(DCF) between these two bands as  it would  
give the time difference as positive or negative lag between these emissions of corresponding bands. 
Unfortunately, we could not do a DCF analysis for this 3C 273 data, as they are quite poorly and irregularly
sampled, so any computed DCF would be only too likely to indicate spurious correlations. 
The optical and UV bands too show correlation between their emissions, again indicating that these two bands 
are emitted from same region, but they are not correlated with X-ray band. Although for most  blazars there is 
 evidence that the X-ray emission is from the high energy tail of the synchrotron emission, which  just 
extends the UV band to the X-ray (Jester et al.\ 2007), but apparently during the period, 2000--2012
this FSRQ apparently did not follow that scenario.  In the 
local environment of 3C 273, there might be two electron populations with two different upper limits to
their Lorentz factors such that
 the one  with higher energetic particles is emitting in both the soft and hard 
X-ray bands while the other one with comparatively lower energetic particles is responsible for optical/UV 
emission.  

Antilockwise loops in the hardness ratio vs 0.3$-$10 keV flux representation, which may also be referred 
to as the spectral index$-$flux representation, usually reveals a hard lag between the two emissions indicating that the 
soft band leads the hard band, yeilds that the source's hard flux increases more than soft flux with increasing 
total flux. This implies that during those periods particle acceleration mechanism was dominating
over synchrotron mechanism for the emission observed in X-ray band.
   
To check the variation of flux over the observing time span we plotted the mean count rate of 
individual observations against modified Julian date. If one discounts some observations at the start 
and end of the totality of the measurements, the source shows a trend of flux decreasing (Fig. 3).

In this paper we have studied X-ray and optical/UV variability properties of the blazar 3C 273 with 
observations taken by the XMM$-$Newton satellite since its launch through the year 2012. We made use of all  
available data in the Public XMM$-$Newton data archive. We use the Epic/pn detector data for our X-ray study 
and the simultaneous OM data to study the source in different optical and UV bands for learn more about
the  physical  processes producing the radiation. Our study of this fascinating source regarding its 
 behaviour in different energy bands over this extended period shows high flux variation 
in long term in all the bands, but the Optical/UV flux variability is much higher than that in X-ray 
bands. Optical and UV variability depends on frequency, and increases with increasing frequency, as found 
in previous studies. But at the same time, the source follows the opposite frequency dependency of 
variability in the case X-ray emissions. During this observed 12 year interval the source shows evidence 
of both synchrotron cooling and particle acceleration mechanisms dominating the emission during different periods. 
No relation 
between X-ray and optical/UV emissions is found, supporting the scenario where two independent particle 
populations are present in its local emitting regions.

\section*{ACKNOWLEDGEMENTS}

We thank the referee for detailed and thoughtful comments which helped us to improve the manuscript. 
We acknowledge the Department of Science \& Technology (DST), India, for supporting this work with
 grants under the Women Scientist scheme-A (WOS--A). This research is based on observations
taken with XMM--Newton, an ESA science mission with instruments and contributions directly funded by
ESA Member Sates and NASA.

{}

\clearpage
\end{document}